\documentclass[runningheads]{llncs}
\usepackage{diagbox}
\usepackage{float} 
\usepackage{booktabs} 
\usepackage{physics}
\usepackage[normalem]{ulem}

\usepackage{verbatim}
\usepackage{filecontents}

\usepackage{pgfplots}
\pgfplotsset{compat=1.8}
\usepackage{pgfplotstable}
\usepgfplotslibrary{groupplots}

\usepackage{marvosym}

\usepackage{balance}       
\usepackage{graphics}      
\usepackage[T1]{fontenc}   
\usepackage{txfonts}
\usepackage{mathptmx}
\usepackage[pdflang={en-US},pdftex]{hyperref}
\usepackage{color}
\usepackage{booktabs}
\usepackage{textcomp}

\usepackage{algorithm}
\usepackage{algorithmic}

\usepackage{microtype}        
\usepackage{ccicons}          

\usepackage{todonotes}

\begin{document}
\title{Quantum-Inspired Keyword Search on Multi-Model Databases}

%
%
\author{Gongsheng Yuan\inst{1,2} \and
Jiaheng Lu\inst{1}\textsuperscript{} \and
Peifeng Su\inst{3}}

\authorrunning{G. Yuan et al.}
%

\institute{Department of Computer Science, University of Helsinki, FI-00014, Helsinki, Finland \\
\email{\{gongsheng.yuan,jiaheng.lu\}@helsinki.fi}\and
School of Information, Renmin University of China, Beijing 100872, China \and
Department of Geosciences and Geography, University of Helsinki, FI-00014, Helsinki, Finland \\
\email{peifeng.su@helsinki.fi}
}
\maketitle              
\begin{abstract}


With the rising applications implemented in different domains, it is inevitable to require databases to adopt corresponding appropriate data models to store and exchange data derived from various sources. To handle these data models in a single platform, the community of databases introduces a multi-model database. And many vendors are improving their products from supporting a single data model to being multi-model databases. Although this brings benefits, spending lots of enthusiasm to master one of the multi-model query languages for exploring a database is unfriendly to most users. Therefore, we study using keyword searches as an alternative way to explore and query multi-model databases. In this paper, we attempt to utilize quantum physics's probabilistic formalism to bring the problem into vector spaces and represent events (e.g., words) as subspaces. Then we employ a density matrix to encapsulate all the information over these subspaces and use density matrices to measure the divergence between query and candidate answers for finding top-\textit{k} the most relevant results. In this process, we propose using pattern mining to identify compounds for improving accuracy and using dimensionality reduction for reducing complexity. Finally, empirical experiments demonstrate the performance superiority of our approaches over the state-of-the-art approaches.

\end{abstract}

\section{Introduction}



In the past decades, due to an explosion of applications with the goal of helping users address various transactions in different domains, there are increasing needs to store and query data produced by these applications efficiently.
As a result, researchers have proposed diverse data models for handling these data, including structured, semi-structured, and graph models. Recently, to manage these data models better, the community of databases introduces an emerging concept, multi-model databases \cite{lu2019multi}, which not only embraces a single and unified platform to manage both well-structured and NoSQL data, but also satisfies the system demands for performance, scalability, and fault tolerance.

\begin{figure*}[t]
\centering
\includegraphics[height=4.2cm]{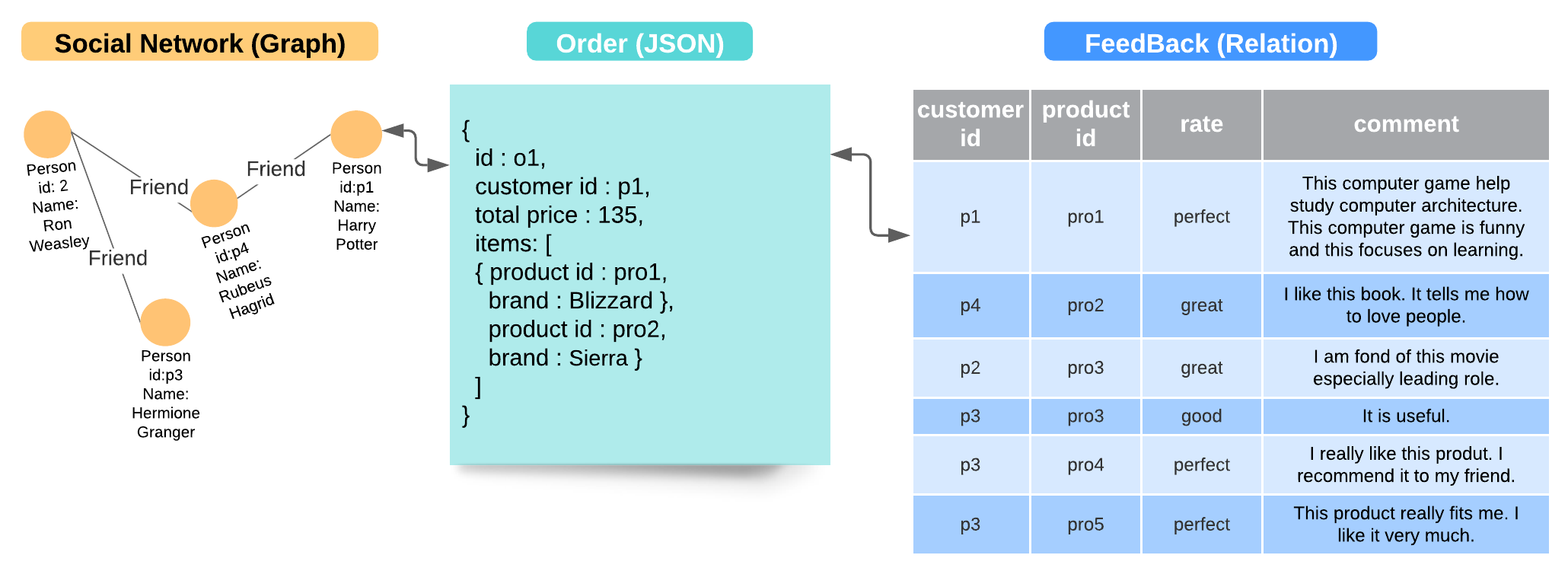}
\caption{An Example of Multi-model Data.}
\label{fig:multimodel}
\end{figure*}

Although multi-model database systems provide a way to handle various data models in a unified platform, users have to learn corresponding specific multi-model query languages to access different databases (e.g., using AQL to access ArangoDB \cite{ArangoDB}, SQL++ for AsterixDB \cite{altwaijry2014asterixdb}, and OrientDB SQL for OrientDB \cite{orientdb2017orientdb}). Moreover, users also need to understand complex and possibly evolving multi-model data schemas as background knowledge for using these query language. This is unfriendly to most users because it usually with a steeper learning curve.
For example, Fig.~\ref{fig:multimodel} depicts multi-model data in social commerce. \textit{Suppose we want to find Rubeus Hagrid's friends who have bought Blizzard and given a perfect rating.} It won't be an easy job to write a multi-model query involving social network (Graph), order (JSON), and feedback (Relation) for novices to achieve this goal.
Therefore, in this paper, we study using keyword searches as an alternative way to explore and query multi-model databases, which does not require users to have strong background knowledge.

After reviewing the literature, we find that most existing works \cite{yu2009keyword} only restrict keyword searches over a specific database supporting a single data model (e.g., relational, XML, and graph databases). Unfortunately, there is a lack of relevant research literature for the issue of performing keyword searches on multi-model databases. However, we think it is a promising research topic that remains a big challenge. This is because a trivial solution, which firstly performs a keyword search on individual data models by conventional methods and then combines results by assembling the previous results, cannot work well. The reason is that it may miss results that consist of multiple models simultaneously.



Facing this challenge, previous researchers used graph methods \cite{li2011effective} to solve it. However, there are several matters needing attention in this method. Firstly, when we use a graph to represent these heterogeneous data, this graph may be vast and complex. So we need to divide this graph into many subgraphs, which is a graph partition problem. And if we perform keyword searches on these graphs to find answers, this means we need to find some subgraphs relating to partial keywords or all as the answer. Therefore, this involves subgraph matching and subgraph relevance ranking problems. And lastly, we need to consider how to take the dependencies among keywords and schema information into consideration when doing keyword searches. We could see all these problems have a significant influence on the final returned results. This means we should be careful to choose the corresponding solutions for these graph problems.


To avoid these graph issues, we start by introducing the ``\textit{quantum-inspired}'' framework into the database community \cite{yuan2020quantum}, in which we utilize the probabilistic formalism of quantum physics to do keyword searches. In quantum probability, the probabilistic space is naturally encapsulated in a vector space. Based on the notion of information need vector space, we could regard the data in multi-model databases as the information (\textit{statements}) collection, define events (e.g., words) as subspaces, use a density matrix (probability distribution) to encapsulate all the information over these subspaces for measuring the relevance of the candidate answers to the user's information need.

For this framework, the idea behind the quantum-inspired approach to disciplines other than physics is that, although macroscopic objects cannot exhibit the quantum properties manifested by particles such as photons, some phenomena can be described by the language or have some features of the phenomena (e.g., superposition) represented by the quantum mechanical framework in physics \cite{van2004geometry}. Therefore, except for the above theoretical method introduction, there are two other reasons underlying this attempt. One is that the similarity between the quantum mechanical framework to predict the values which can only be observed in conditions of uncertainty \cite{van2004geometry} and the decision about the relevance of the content of a text to an information need is subject to uncertainty \cite{melucci2013deriving}. Another one is that increasing works support the notion that quantum-like phenomena exist in human natural language and text, cognition and decision making \cite{song2010quantum}, all related to the critical features of keyword searches.

Besides, the pioneering work \cite{van2004geometry} formalized quantum theory as a formal language to describe the objects and processes in information retrieval. Based on this idea, we use this mathematical language to describe relational, JSON, and graph data in the database community as information collection. Next, we transform keyword searches from a querying-matching work into a calculating-ranking task over this information collection and return the most relevant top-$k$ results. And we take the possible relevance among input query keywords and database schema information into consideration, which helps the framework understand the user's goal better. Now, we summarize our contributions as follows:

\begin{enumerate}
\item Based on quantum theory, we attempt to use a quantum-inspired framework to do keyword searches on multi-model databases, utilizing quantum physics' probabilistic formalism to bring the problem into information need vector space. We want to take advantage of the ability of quantum-inspired framework in capturing potential semantics of candidate answers and query keywords for improving query accuracy.


\item In this process, we introduce the co-location concept to identify compounds for enhancing the topic of statements. We want to use it to improve query performance (precision, recall, and F-measure).

\item By analyzing the existing quantum-inspired method, we propose constructing a query density vector instead of a density matrix to reduce the framework's complexity. And we present an algorithm to perform keyword searches on multi-model databases.

\item Finally, we perform extensive empirical experiments that demonstrate our approaches' overall performance is better than the state-of-the-art approaches. The F-measure is at least nearly doubled.
\end{enumerate}



\section{Preliminaries}
For the sake of simplicity, we assume the vector space is $\mathbb{R}^{n}$ in this paper. A unit vector \(\vec{u}\in\mathbb{R}^{n}\) is defined as $\ket{u}$  and termed as \textit{ket} on the basis of Dirac notation. 
Its conjugate transpose $\vec{u}^{\,H} = \vec{u}^{\,T}$ is written as $\bra{u}$ and called \textit{bra}. 
The inner product between two vectors is denoted as $\braket{u}{v} = \sum_{i=1}^{n}u_i v_i$.
The outer product between two vectors is denoted as $\ket{u}\bra{v}$ and called \textit{dyad}.
When the vector has a unitary length (i.e., $\|\vec{u}\|_2 = 1$), a special operator can be defined as the outer product between two vectors. We call this operator \textit{projector}. To explain this, suppose $\ket{u}$ is a vector; the projector corresponding to this vector is a dyad written as $\ket{u}\bra{u}$. For example, if there is $\ket{u_1} = (1, 0)^T$, the projector corresponding to this ket is transforming it into
$
\ket{u_1}\bra{u_1} = 
\begin{pmatrix}
1 & 0 \\
0 & 0
\end{pmatrix}.
$
Due to the projector, $\ket{u}$ could be mapped to the generalized probability space. In this space, each rank-one dyad $\ket{u}\bra{u}$ can represent a quantum \textit{elementary event} and each dyad $\ket{\kappa} \bra{\kappa}$ represent a \textit{superposition event}, where $\ket{\kappa} = \sum_{i = 1} ^p \sigma _ i \ket{u_{i}}$, the coefficients $\sigma_i \in \mathbb{R}$ satisfy $\sum_i \sigma_i ^2 = 1$. And density matrices $\rho$ are interpreted as generalized probability distributions over the set of dyads when its dimension greater than 2 according to Gleason's Theorem \cite{gleason1957measures}. A real density matrix $\rho$ is a positive semidefinite ($\rho \geq 0$) Hermitian matrix ($\rho$ = $\rho^H$ = $\rho^T$) and has trace 1 ($Tr(\rho) = 1$). It assigns a generalized probability to each dyad $\ket{u} \bra{u}$, whose formula is:
\begin{equation}
    \mu_\rho(\ket{u} \bra{u}) = Tr(\rho \ket{u} \bra{u}).
    \label{equ:trace}
\end{equation}
For example,
$
\rho_1 = 
\begin{pmatrix}
0.75 & 0 \\
0 & 0.25
\end{pmatrix}, 
\ 
\rho_2 =  
\begin{pmatrix}
0.5 & 0.5 \\
0.5 & 0.5
\end{pmatrix}
$, density matrix $\rho_2$ assigns a probability value $Tr(\rho_2 \ket{u_1} \bra{u_1}) = 0.5$ to the event $\ket{u_1} \bra{u_1}$. If $\rho$ is unknown, we could utilize the \textit{Maximum Likelihood} (MaxLik) estimation to get it. Finally, through the value of negative Von-Neumann Divergence (VND) between $\rho_q$ (query) and $\rho_c$ (candidate answer), we could get their difference. Its formalization is:
\begin{equation}
-\Delta_{VN}(\rho_q||\rho_c) \overset{rank}{=} \sum_i \lambda_{q_i} \sum_j(\log \lambda_{c_j} \bra{q_i}\ket{c_j} ^2).
\label{equ:VND}
\end{equation}

\begin{figure*}[t]
    \centering
    \includegraphics[height=3.7cm,width=11cm]{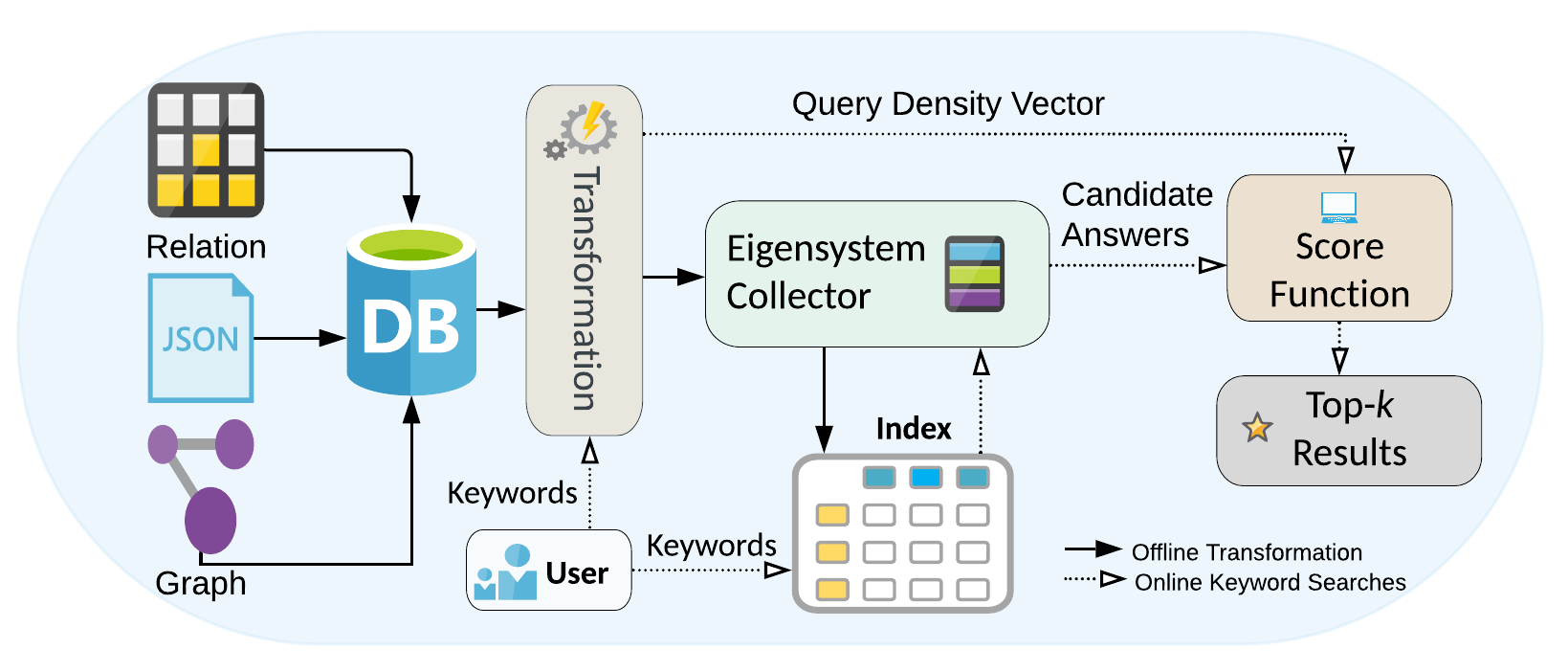}
    \caption{The Framework of Quantum-Inspired Keyword Search}
    \label{fig:framework}
\end{figure*}

\section{The framework of the Quantum-inspired Keyword search}
An overview of this entire framework is shown in Fig.~\ref{fig:framework}, which has two functional modules. One is offline transformation. We will use quantum language to represent heterogeneous data in a uniform format. Another one is online keyword searches with generalized probabilities for finding the most relevant results to answer keyword searches.

\subsection{Transformation}

Here we consider three kinds of data. They are relational, JSON, and graph data. For the \textbf{relational} data model, it is natural to think of a tuple with schema in tables as a \textit{statement} (a piece of information).
To avoid information explosion (caused by information fragmentation), we observe the \textbf{JSON} file in a coarse granularity and treat each object in JSON as an integral \textit{statement}.
For each node in the \textbf{graph} data, we put information about all of its neighbor nodes in the same \textit{statement}, including itself information. In this way, it can keep complete neighbor relationships and node information.

Only considering these discrete data is not enough. We need to take join into consideration to get complete information in our framework. For this problem, there are many methods. For example, we could mine the relationship among different data models with \cite{Bergamaschi2011KeywordSO}, then join them. Or we could choose meaningfully related domains to do equi-join operations based on expert knowledge, which is a quite straightforward and easy way for finding meaningful joins and cutting down the search space by avoiding generating a lot of useless intermediate results. Since this part is not the focus of this paper, we will use the later one to continue our work.

Now we have gotten a statement collection. Next, we use the mathematical language of quantum theory to represent these statements. To achieve this goal, we first use an elementary event to represent a single word and use a superposition event to represent a compound. In this process, to construct proper superposition events to enhance the topic of statement for improving query accuracy (i.e., choose appropriate compounds $\kappa$ from a given statement and determine the value of each coefficient $\sigma_i$ in $\kappa$), we introduce a special spatial pattern concept called \textit{co-location pattern} \cite{huang2004discovering}. After this process, we could get an event set $\mathcal{P}_{st}$ for each statement. Then, we learn a density matrix from each event set $\mathcal{P}_{st}$ with MaxLik estimation and use this density matrix to represent the corresponding statement. Here, we use a $R$$\rho$$R$ algorithm, which is an iterative schema of MaxLik outlined in the \cite{lvovsky2004iterative}. And this iterative schema has been used in the \cite{sordoni2013modeling} for getting density matrices. Finally, we could use VND to measure the difference between different statements by their density matrices and use VND' values to rank them.

To get an event set $\mathcal{P}_{st}$ for each statement, we introduce a special spatial pattern concept called \textit{co-location pattern} \cite{huang2004discovering} to help construct compounds $\kappa$. Next, we will start by redefining some concepts in the field of co-location to adapt our problem for identifying compounds. When a word set $c = \{w_1, ..., w_p\}, \norm{c} = p$, appears in any order in a fixed-window of given length $L$, we say there is a relationship $R$ among these words. For any word set $c$, if $c$ satisfies $R$ in a statement, we call $c$ co-location compound.

\begin{definition} \textbf{Participation ratio (PR)} $PR(c,w_i)$ represents the participation ratio of word $w_i$ in a co-location compound $c = \{w_1, ..., w_p\}$, i.e.
  \begin{equation}
    PR(c,w_i)=  \frac{T(c)}{T(w_i)} \quad w_i \in c,
    \label{equ:PR}
  \end{equation}
where $T(c)$ is how many times the $c$ is observed together in the statement , $T(w_i)$ is the number of word $w_i$ in the whole statement.
\end{definition}
\begin{definition} \textbf{Participation index (PI)} $PI(c)$ of a co-location compound $c$ is defined as: 
  \begin{equation}
    PI(c)=\min_{i=1}^p \{PR(c,w_i)\}.
    \label{equ:PI}
  \end{equation}
\end{definition}

\begin{table}[!tbp]
  \caption{Example about identifying compound $\kappa$}
  \label{tab:co-location}
  \begin{tabular}{m{4.9cm}m{3.1cm}m{1cm}m{1.3cm}m{1cm}}
    \toprule
     \multicolumn{5}{c}{ $\mathcal{W}_{st}$ = \{\textit{This computer game help study computer architecture} }\\
     \multicolumn{5}{c}{\textit{this computer game is funny and this focuses on learning.} \} }
    \\    \midrule
    $c_1$ = \{ \textit{computer, game} \}
    &
    PR( $c_1$, computer ) \newline
    PR( $c_1$, game )
    &
    $\frac{2}{3}$ \newline
    $\frac{2}{2}$ 
    &
    PI( $c_1$ )
    &
    $\frac{2}{3}$
    \\    \midrule  
    $c_2$ = \{ \textit{game, architecture} \}
    &
    PR( $c_2$, game ) \newline
    PR( $c_2$, architecture )
    &
    $\frac{1}{2}$ \newline
    $\frac{1}{1}$
    &
    PI( $c_2$ )
    &
    $\frac{1}{2}$
    \\    \midrule  
    $c_3$ = \{ \textit{computer, architecture} \}
    &
    PR( $c_3$, computer ) \newline
    PR( $c_3$, architecture )
    &
    $\frac{1}{3}$ \newline
    $\frac{1}{1}$
    &
    PI( $c_3$ )
    &
    $\frac{1}{3}$
    \\    \midrule  
    $c_4$ = \{ \textit{computer, game, architecture} \} 
    &
    PR( $c_4$, computer ) \newline
    PR( $c_4$, game ) \newline
    PR( $c_4$, architecture )
    &
    $\frac{1}{3}$ \newline
    $\frac{1}{2}$ \newline
    $\frac{1}{1}$
    &
    PI( $c_4$ )
    &
    $\frac{1}{3}$
    \\    \midrule  
     \multicolumn{5}{c} 
     {
        PI( $c_1$ ) $\geq min\_threshold$ = 0.6, $c_1$ is a compound $\kappa$
    }
  \\ \bottomrule
\end{tabular}
\end{table}

Given a minimum threshold $min\_threshold$, a co-location compound $c$ is a compound $\kappa$ if and only if $PI(c) \geq min\_threshold$. It is obvious that when $min\_threshold$ = 0, all the co-location compounds are compounds. Based on the method of estimating whether $c$ is a compound $\kappa$, with each appearance of word $w_i \in c$ in the statement, there has been a great effect on calculating value of $PR$ which determines whether $PI(c)$ is greater than or equal to $min\_threshold$. This is to say each appearance of word $w_i$ plays an important role in the statement for expressing information. Therefore, when we set a value for the coefficient $\sigma_i$ of $\ket{w_i}$ in $\kappa$ , we need to take the above situation into consideration. A natural way is to set $\sigma_i^2$ = $T(w_i)/ \sum_{j=1} ^p T(w_j)$ for each $w_i$ in a compound $\kappa = \{w_1, ..., w_p\}, 1 \leq $i$ \leq $ $p$, where $T(w_i)$ is the number of times that word $w_i$ occurs in this statement. We call  it \textit{co-location weight}.

For example, we assume ``This computer game help study ...'' is a statement $\mathcal{W}_{st}$, which is in the table FeedBack of Fig.~\ref{fig:multimodel}. Given $c_1$, $c_2$, $c_3$, and $c_4$ (see Table~\ref{tab:co-location}), if $min\_threshold$ = 0.6, thus $c_1$ is a compound. 
This helps our framework to have a better understanding of this statement whose theme is about the computer game, not computer architecture. Meanwhile, with co-location weight we could set $\sigma_{computer} ^2$ = $3/5$, $\sigma_{game}^2$ = $2/5$ for the $\kappa$ = \{\textit{computer, game}\}.
According to the related state-of-the-art work \cite{Bendersky:2012:MHT:2348283.2348408}, the fixed-window of Length $L$ could be set to $l||c||$. And it provides a robust pool for $l$. Here, we select $l$ = 1 from the robust pool. Because if one decides to increase $l$, more inaccurate relations will be detected, and performance will deteriorate.

Next, we could use an iterative schema of $MaxLik_{matrix}$ \cite{sordoni2013modeling} to learn a density matrix from each event set $\mathcal{P}_{st}$. However, unfortunately, both dyads and density matrices are $n \times n$ matrices, i.e., their dimensions depend on $n$, the size of word space $\mathcal{V}$. When $n$ is bigger, the cost of calculating is higher, especially in the field of databases. Therefore, we need to find a way to reduce complexity.

According to the Equation~(\ref{equ:trace}), the probability of quantum event $\Pi$ is $Tr(\rho \Pi)$. Through $Tr(\rho \Pi)$, we could get the dot product of two vectors, i.e., 
\begin{equation}
Tr(\rho \Pi) = \vec{\rho}.\vec{\Pi}^2,
\label{equ:decomposing}
\end{equation}
where $\vec{\rho}$ = ($\beta_1$, $\beta_2$, $\ldots$, $\beta_n$), we call it \textit{density vector}. If we assume $\vec{\Pi}$ = ($(\sum_{i=1}^{n}u_{i1} v_{i1})$, $(\sum_{i=1}^{n}u_{i1} v_{i2})$, $\ldots$, $(\sum_{i=1}^{n}u_{i1} v_{in})$), then $\vec{\Pi}^2$ = ($(\sum_{i=1}^{n}u_{i1} v_{i1})^2$, $(\sum_{i=1}^{n}u_{i1} v_{i2})^2$, $\ldots$, $(\sum_{i=1}^{n}u_{i1} v_{in})^2$).

\begin{definition}
 \textbf{Density vector}
    A density vector is a real vector 
    $\vec{\rho}$ = $\begin{pmatrix} \beta_1, \beta_2, \ldots, \beta_n \end{pmatrix}$, noted by $\bra{\rho}$, where $\beta_i \geq 0$, and $\beta_1 + \beta_2 + \ldots + \beta_n = 1$. 
\end{definition}

For example, density vector $\bra{\rho_3}$ = (0.2, 0.1, 0.7). Next, we will use a density vector instead of a density matrix in the rest of this paper. This is because it has three advantages.
\begin{itemize}
    \item Firstly, density vectors have a more concise way of calculating quantum probabilities than density matrices' (comparing Equation \ref{equ:trace} with Equation \ref{equ:DensityVecProb});
    \item Secondly, when one calculates VND between two density vectors, it is faster to get the value of VND than density matrices do (comparing Equation \ref{equ:VND} with Equation \ref{equ:VecVND});
    \item Thirdly, comparing with density matrices, learning a query density vector from the input keywords is easier.
\end{itemize}

Based on the generated density matrices, we could get $val$ (the list of all eigenvalues) and $vec$ (the list of eigenvectors, each eigenvector corresponding to its own eigenvalue) through eigendecomposition of density matrices. To simplify our framework further, we use the Principal Component Analysis (PCA) method to take the first $h$ largest eigenvalues and require that the sum of these values is greater than or equal to a $threshold$. Based on the research concerning PCA \cite{holland2008principal}, we could set $threshold$ = 85\%, which is enough for our work. This is because if one increases $threshold$, a higher $threshold$ may have a higher chance to cause the increase of the number of $val_i$ instead of precision of framework and deteriorate the performance of the framework. Then, we store $h$ eigenvalues and corresponding eigenvectors $vec_i$ (termed as $\{val, vec\}$) into the set of density systems, $S_{denSys}$ (the component, "Eigensystem Collector" in Fig.\ref{fig:framework}).

We have now transformed all kinds of data into eigenvalues (density vectors) and eigenvectors (mapping directions) and converted querying-matching problem into calculating-ranking. In this way, we eliminate heterogeneous data and convert the querying-matching problem into calculating-ranking. Next, we will consider how to do keyword searches on these density vectors online.


\subsection{Online Keyword Search Query}
In this module, when users submit their queries, there are two data flows about queries.
\begin{itemize}
    \item The first one is from users to "Transformation", which will construct query density vectors from input keywords. This step will consider the relationship between input keywords instead of treating keywords as mutually independent entities.
    \item Another one is from users to index structure for getting candidate answers, which will reduce the query scope through our index structure (inverted list).
\end{itemize}

In the above data flows, it involves constructing query density vectors. Before doing it, we propose the new representation for single words and compounds in query keywords. Through analyzing Equation~(\ref{equ:decomposing}), we could represent each word $w_i \in \mathcal{V}$ by $\ket{e_i}$, where $\ket{e_i}$, the standard basis vector, is an one-hot vector. Based on this new representation, we present compound by $\ket{\kappa} = \sum_{i = 1}^p \sigma _ i \ket{e_{w_i}}$, where $\kappa = \{ w_1, ..., w_p\}$, the coefficients $\sigma_i \in \mathbb{R}$ satisfy $\sum_{i=1}^p \sigma_i^2$ = 1 to guarantee the proper normalization of $\ket{\kappa}$.

For example, Considering $n = 3$ and $\mathcal{V}$ = \{$computer$, $science$, $department$\}, if $\kappa_{cs}$ = \{$computer$, $science$\} and $\ket{\kappa_{cs}}$ = $\sqrt{2/3}\ket{e_c}$ + $\sqrt{1/3}\ket{e_s}$, then we could get:
\begin{displaymath}
\begin{aligned}
    \ket{e_{computer}} = \begin{pmatrix} 1 \\ 0 \\  0\end{pmatrix},
    &\quad
    \ket{e_{science}} = \begin{pmatrix} 0 \\ 1 \\ 0\end{pmatrix},
    \quad
    \ket{\kappa_{cs}} = 
    \sqrt{\frac{2}{3}}
    \begin{pmatrix} 1 \\ 0  \\ 0 \end{pmatrix}
    +
    \sqrt{\frac{1}{3}}
    \begin{pmatrix} 0 \\ 1 \\ 0 \end{pmatrix} = 
    \begin{pmatrix} \sqrt{\frac{2}{3}} \\ \sqrt{\frac{1}{3}}  \\ 0 \end{pmatrix}.
\end{aligned}
\end{displaymath}


Next, we still use the iterative schema of \textit{Maximum Likelihood} (MaxLik) estimation for getting the query density vector. We would also use the following formula to get the quantum probability of event $\ket{\Pi}$ at $MaxLik$ estimation.
\begin{equation}
    Probability_{\Pi} = \braket{\rho}{\Pi^2}.
    \label{equ:DensityVecProb}
\end{equation}

\begin{algorithm}[t]
\begin{algorithmic}[1]
    \renewcommand{\algorithmicrequire}{\textbf{Input:}}
    \renewcommand{\algorithmicensure}{\textbf{Output:}}
    \REQUIRE Input keywords $K$ = \{$w_1$, \ldots, $w_t$\}
    \ENSURE  the top-$k$ most relevant statements about queries
    \STATE  \{\textit{$K_{schema}$, $K_{non\_schema}$}\} = $Classification$($K$)
    \STATE  $C_{schema}$ $\gets$ $\Phi$\;
    \FOR {each $w_i \in K_{schema}$} 
         \STATE $C_{schema}$ $\gets$ $C_{schema}$ $\cup$ ($C_{w_i}$ $\subset$ $S_{denSys}$)
    \ENDFOR

    \FOR {all $w_j$ $\in$ \ $K_{non\_schema}$} 
        \STATE  $C_{non\_schema}$ $\gets$ $C_{w_1}$\ $\cap$ , \ldots, $\cap$\ $C_{w_j}$
    \ENDFOR
    
    \STATE  $C_{denSys}$ $\gets$ $C_{non\_schema}$ - $C_{schema}$
    \STATE  $S_{Result}$ $\gets$ $\Phi$
    \FOR{each $\{val, vec\}_{st}$ $\in$ $C_{denSys}$}
         \STATE     $\mathcal{P}_{q}$ $\gets$ $\Phi$
         \FOR {each $w_i \in K$} 
            \STATE Get event $\vec{\Pi}_{w_i}$ through rotating the $\ket{e_i}$ $(w_i)$ into a new coordinate by $vec_{st}$
            \STATE     $\mathcal{P}_{q}$ $\gets$ $\mathcal{P}_{q}$ $\cup$ $\vec{\Pi}_{w_i}$  
        \ENDFOR
        
        \FOR {each $c\ in \ K$} 
            \IF{$PI(c)$ $\geq$  \textit{min\_{threshold}}} 
                \STATE $\mathcal{P}_{q}$ $\gets$ $\mathcal{P}_{q}$ $\cup$ $\ket{\kappa_c}$ 
            \ENDIF
        \ENDFOR
    
        \STATE Learn a query density vector $\bra{\rho}_q$ from $\mathcal{P}_{q}$ by $MaxLik$

        \STATE  $score_{st}$ = $ScoreFunction$($\bra{\rho}_q$, $val_{st}$)
        \STATE  $S_{Result}$ $\gets$ $S_{Result}$ $\cup$ $score_{st}$
    \ENDFOR
    \STATE  Sort $S_{Result}$ and return top-$k$ results
\end{algorithmic}
\caption{Answer Keyword Searches with Density Vectors}
\label{alg:approach2}
\end{algorithm}

Finally, we propose Algorithm~\ref{alg:approach2} to do online keyword searches, in which we use line 1 - 9 to get candidate density system set $C_{denSys}$ through the index. Then we use them to help get corresponding query density vectors and regard them as the score function (VND) inputs to answer queries (line 10 - 26).

Firstly, we could classify the input keywords into two groups according to whether or not $w_i$ is relevant to the database schema. For the group $K_{schema}$, we use the union to get a candidate density system set $C_{schema}$ in line 4, where $C_{w_i}$ can be gotten by the inverted list, which is relevant to the input word $w_i$. For the $K_{non\_schema}$, we use intersection to get $C_{non\_schema}$. Then we could get the candidate density system set $C_{denSys}$ through difference (line 9). We use these operations to take advantage of the schema for helping users explore more potential possibilities about answers.

Line 10 - 26 are going to get top-$k$ results.
Firstly, we use every mapping direction $vec_{st}$ of $\{val, vec\}_{st}$ in the candidate density systems $C_{denSys}$ to transform original events into new representations so that the algorithm could learn a query density vector from these new events. The above transforming could guarantee that each constructed query density vector and corresponding density vector $val_{st}$ in one coordinate system.
Then we could calculate the divergence between $\bra{\rho}_q$ and $\bra{\rho}_{st}$ ($val_{st}$) by Equation \ref{equ:VecVND} in line 23.


\begin{equation}
\begin{aligned}
-\Delta_{VN}(\bra{\rho}_q||\bra{\rho}_{st})
&\overset{rank}{=} \sum_i \beta_{q_i} \log \beta_{st_i}
, \ where \ \beta_{q_i} \in \bra{\rho}_q, \beta_{st_i} \in \bra{\rho}_{st}.
\end{aligned}
\label{equ:VecVND}
\end{equation}

Considering the previous problem "\textit{Suppose we want to find Rubeus Hagrid's friends who have bought Blizzard and given a perfect rating.}", we take \{Rubeus Hagrid friends Blizzard perfect\} as Algorithm~\ref{alg:approach2} inputs and perform keyword searches on the Fig.~\ref{fig:multimodel}. And we could get the result in which a person named Harry Potter bought Blizzard and gave a perfect rating, and he is Rubeus Hagrid's friend. The original result is ``\textit{social network person id p1 name harry potter friend person id p4 name rubeus hagrid order id o1 custom id p1 total price 135 item product id pro1 brand blizzard feedback rate perfect comment this computer game help study computer architecture this computer game is funny and this focuses on learning}''. And its score is -2.665.






\section{Experiment}

\subsection{Data Sets}
\begin{table}[!tbp]
\caption{The number of records/objects in different data models}
\centering
\label{tab:dataset}
\begin{tabular}{p{2.2cm}<{\centering}|p{2.2cm}<{\centering}|p{2.2cm}<{\centering}|p{2.2cm}<{\centering}|p{2.2cm}<{\centering}}
\hline
    &   Relational  &   JSON    &   Graph-entity    &   Graph-relation \\ \hline
UniBench    &   150 000 &   142 257 &   9 949   &   375 620  \\ \hline
IMDB        &   494 295 &   84 309  &   113 858 &   833 178 \\\hline
DBLP    &    1 182 391  &   435 469 &   512 768 &   3 492 502    \\ \hline
\end{tabular}
\end{table}
\begin{table}[!tbp]
\centering
\caption{Queries employed in the experiments}
\label{tab:queries}
\begin{tabular}{ll}
\toprule
\multicolumn{1}{l}{ID}  \quad  &   \multicolumn{1}{l}{Queries}     \\ \midrule
\multicolumn{2}{l}{(a) Queries on DBLP}     \\
$Q_1$  &  Soni Darmawan friends         \\ 
$Q_2$  &  Gerd Hoff friends' rank 1 paper   \\
$Q_3$  &  Slawomir Zadrozny rank 1 paper\\
$Q_4$  &  Phebe Vayanos phdthesis paper        \\
$Q_5$  &  neural sequence model  \\ 
$Q_6$  &  Brian Peach 2019 papers   \\
$Q_7$  &  Performance of D2D underlay and overlay for multi-class elastic traffic. authors\\
$Q_8$  &  Carmen Heine rank Modell zur Produktion von Online-Hilfen.     \\
$Q_9$  &  Exploring DSCP modification pathologies in the Internet.   \\
$Q_{10}$     &  The papers of Frank Niessink          \\
\\
\multicolumn{2}{l}{(b) Queries on UniBench}   \\ 
$Q_{11}$  &  Abdul Rahman Budjana friends BURRDA feedback perfect    \\
$Q_{12}$  &  Shmaryahu Alhouthi order Li-Ning Powertec Fitness Roman Chair            \\
$Q_{13}$  &  Kamel Abderrahmane Topeak Dual Touch Bike Storage Stand       \\
$Q_{14}$  &  Alexandru Bittman whether has friend Ivan Popov \\ 
$Q_{15}$  & Mohammad Ali Forouhar Oakley Radar Path Sunglasses\\
$Q_{16}$  &  Soft Air Thompson 1928 AEG Airsoft Gun and Genuine Italian Officer's Wool Blanket\\
$Q_{17}$  &  Roberto Castillo Total Gym XLS Trainer and Reebok \\
$Q_{18}$  &  Who Kettler, Volkl and Zero Tolerance Combat Folding Knife \\
$Q_{19}$  &  Francois Nath Nemo Cosmo Air with Pillowtop Sleeping Pad \\
$Q_{20}$  &  Hernaldo Zuniga Advanced Elements AdvancedFrame Expedition Kayak and TRYMAX\\
\\
\multicolumn{2}{l}{(c) Queries on IMDB}   \\ 
$Q_{21}$  & Lock, Stock and Two Smoking Barrels actors \\
$Q_{22}$  & Forrest Gump \\
$Q_{23}$  & The title and imdbVote of films of Bruce Willis \\
$Q_{24}$  & The films of director Robert Zemeckis \\ 
$Q_{25}$  & films in 1997 Genre Action, Adventure, Family  \\
$Q_{26}$  & The Legend of 1900 awards\\
$Q_{27}$  & Scent of a Woman imdbRating\\
$Q_{28}$  & The film of Dustin Hoffman with Tom Cruise \\
$Q_{29}$  & Morgan Freeman friends\\
$Q_{30}$  & Aamir Khan films \\
\bottomrule
\end{tabular}
\end{table}

We use synthetic data (UniBench) and real data (IMDB and DBLP) to evaluate our approaches. The statistics about them are listed in Table~\ref{tab:dataset}.

UniBench~\cite{zhang2018unibench} is a multi-model benchmark, including data of relational, JSON, and graph models. It simulates a social commerce scenario that combines the social network with the E-commerce context. The relational model includes the structured feedback information; The JSON model contains the semi-structured orders; The social network is modeled as a graph, which contains one entity and one relation, i.e., customer, and person knows person. These also have correlations across the data models. For instance, the customer makes transactions (Graph correlates with JSON).

The IMDB dataset is crawled from website\footnote{https://www.omdbapi.com/} by OMDB API. Through extracting inner relationships and potential information, we generate several data models to represent the original data. The relational data includes performing information and rating information, which are stored in different tables; The JSON model is made up of film information (e.g., imdbID, title, and year); The graph is about cooperative information, where two actors would be linked together if they have ever worked for the same movie.

The DBLP\footnote{https://dblp.uni-trier.de/xml/} data consists of bibliography records in computer science. Each record in DBLP is associated with several attributes such as authors, year, and title. The raw data is in XML format. Here we describe it in three data models. The publication records are presented in relational data, including author id, paper id, and the author's rank in the author list. A subset of the papers' attributes (e.g., paper id, key, and title) are represented in JSON. We also construct a co-authorship (friend) graph where two authors are connected if they publish at least one paper together.

%
\begin{table}[!htbp]
\caption{Presion, Recall, F-measure on DBLP}
\label{tab:prfDBLP}
\centering
\pgfplotstabletypeset[
	column type=l,
	font=\small,
	every head row/.style={
		before row={%
			\toprule
			& \multicolumn{7}{c}{AKSDV} & \multicolumn{1}{c}{EASE} \\
		},
		after row=\midrule,
	},
	every last row/.style={after row=\bottomrule},
	columns/firstColumn/.style = {column type=c, column name = \textit{min\_threshold}},
	columns/oneApproach0/.style = {column type=p{1.2cm}<{\centering}, column name=0},
	columns/oneApproach2/.style = {column type=p{1.2cm}<{\centering}, column name=0.2},
	columns/oneApproach4/.style  ={column type=p{1.2cm}<{\centering}, column name=0.4},
	columns/oneApproach6/.style = {column type=p{1.2cm}<{\centering}, column name=0.6},
	columns/oneApproach8/.style  = {column type=p{1.2cm}<{\centering}, column name=0.8},
	columns/oneApproach10/.style  ={column type=p{1.2cm}<{\centering}, column name=1.0},
	columns/oneApproachNon/.style  = {column type=p{1.2cm}<{\centering}, column name=non},
	columns/ease/.style = {column type=|p{1.2cm}<{\centering}, column name=},
	col sep=&,row sep=\\,
	string type,
]{
firstColumn & oneApproach0  & oneApproach2 & oneApproach4  & oneApproach6 & oneApproach8  & oneApproach10 & oneApproachNon  & ease\\
Precision & 0.830 & 0.847 & 0.847 & 0.847 & 0.847 & 0.847 & 0.797 & 0.090 \\
Recall    & 0.867 & 0.917 & 0.917 & 0.917 & 0.917 & 0.917 & 0.817 & 0.500 \\
F-measure & 0.834 & 0.861 & 0.861 & 0.861 & 0.861 & 0.861 & 0.794 & 0.141\\
}
\end{table}

\subsection{Queries and Answers}
In the experiments, three groups of keyword queries, as shown in Table~\ref{tab:queries}, are proposed by a few people randomly to evaluate our methods on the DBLP, UniBench, and IMDB datasets, respectively. Each keyword query involves one or more than one data model to test the ability of our methods in capturing potential semantics of keywords and search accuracy. Besides, the corresponding AQL query for each $Q_i$ is issued in ArangoDB, and the output answers are used to evaluate the results produced by the algorithm, Answer Keyword Searches with Density Vectors (AKSDV), and EASE \cite{li2011effective}. EASE models heterogeneous data as graphs and aims at finding r-radius Steiner graphs as query results. In this method, each returned Steiner graph includes at least two keywords.

The experiments are implemented in Java except for eigendecomposition by Matlab (offline work). The experiments are run on a desktop PC with an Intel(R) Core(TM) i5-6500 CPU of 3.19 GHz and 16GB RAM. Note that all operations are done in memory, and the standard NLP pre-processing such as dropping the stop words and stemming are conducted in advance. In the experiments, we measure the precision, recall, and F-measure for the top-20 returned results.

%
\begin{table}[!tbp]
\caption{Presion, Recall, F-measure on UniBench}
\label{tab:prfUniBench}
\centering
\pgfplotstabletypeset[
	column type=l,
	font=\small,
	every head row/.style={
		before row={%
			\toprule
			& \multicolumn{7}{c}{AKSDV} & \multicolumn{1}{c}{EASE}\\
		},
		after row=\midrule,
	},
	every last row/.style={after row=\bottomrule},
	columns/firstColumn/.style = {column type=c, column name = \textit{min\_threshold}},
	columns/oneApproach0/.style = {column type=p{1.2cm}<{\centering}, column name=0},
	columns/oneApproach2/.style = {column type=p{1.2cm}<{\centering}, column name=0.2},
	columns/oneApproach4/.style  ={column type=p{1.2cm}<{\centering}, column name=0.4},
	columns/oneApproach6/.style = {column type=p{1.2cm}<{\centering}, column name=0.6},
	columns/oneApproach8/.style  = {column type=p{1.2cm}<{\centering}, column name=0.8},
	columns/oneApproach10/.style  ={column type=p{1.2cm}<{\centering}, column name=1.0},
	columns/oneApproachNon/.style  = {column type=p{1.2cm}<{\centering}, column name=non},
	columns/ease/.style = {column type=|p{1.2cm}<{\centering}, column name=},
	col sep=&,row sep=\\,
	string type,
]{
firstColumn & oneApproach0  & oneApproach2 & oneApproach4  & oneApproach6 & oneApproach8  & oneApproach10 & oneApproachNon & ease\\
Precision & 0.902 & 0.902 & 0.917 & 0.922 & 0.922 & 0.922 & 0.897 & 0.220  \\
Recall    & 0.883 & 0.883 & 0.885 & 0.886 & 0.886 & 0.886 & 0.882 & 0.136 \\
F-measure & 0.844 & 0.844 & 0.848 & 0.850 & 0.850 & 0.850 & 0.842 & 0.061\\
}
\end{table}
\begin{table}[!tbp]
\caption{Presion, Recall, F-measure on IMDB}
\label{tab:prfIMDB}
\centering
\pgfplotstabletypeset[
	column type=l,
	font=\small,
	every head row/.style={
		before row={%
			\toprule
			& \multicolumn{7}{c}{AKSDV} & \multicolumn{1}{c}{EASE}\\
		},
		after row=\midrule,
	},
	every last row/.style={after row=\bottomrule},
	columns/firstColumn/.style = {column type=c, column name = \textit{min\_threshold}},
	columns/oneApproach0/.style = {column type=p{1.2cm}<{\centering}, column name=0},
	columns/oneApproach2/.style = {column type=p{1.2cm}<{\centering}, column name=0.2},
	columns/oneApproach4/.style  ={column type=p{1.2cm}<{\centering}, column name=0.4},
	columns/oneApproach6/.style = {column type=p{1.2cm}<{\centering}, column name=0.6},
	columns/oneApproach8/.style  = {column type=p{1.2cm}<{\centering}, column name=0.8},
	columns/oneApproach10/.style  ={column type=p{1.2cm}<{\centering}, column name=1.0},
	columns/oneApproachNon/.style  = {column type=p{1.2cm}<{\centering}, column name=non},
	columns/ease/.style = {column type=|p{1.2cm}<{\centering}, column name=},
	col sep=&,row sep=\\,
	string type,
]{
firstColumn & oneApproach0  & oneApproach2 & oneApproach4  & oneApproach6 & oneApproach8  & oneApproach10 & oneApproachNon & ease\\
Precision & 0.753 & 0.753 & 0.753 & 0.758 & 0.758 & 0.758 & 0.758 & 0.548 \\
Recall    & 0.782 & 0.782 & 0.782 & 0.784 & 0.784 & 0.784 & 0.784 & 0.466 \\
F-measure & 0.657 & 0.657 & 0.657 & 0.661 & 0.661 & 0.661 & 0.661 & 0.269\\
}
\end{table}

\subsection{Results Analysis}
Table~\ref{tab:prfDBLP} shows the comparison of the average precision, recall, and F-measure of AKSDV with EASE's. This comparison result demonstrates that our proposed methods outperform EASE on the DBLP dataset. Table~\ref{tab:prfUniBench} and Table~\ref{tab:prfIMDB} show that the performance of AKSDV also outperforms EASE's on the UniBench and IDMB dataset. And the F-measure of AKSDV is at least nearly twice EASE's on these datasets. These high accuracy values show that our framework could understand the potential semantics underlying the statements and get the most relevant statements about queries.


For example, $Q_9$ wants to find all information about the paper ``Exploring DSCP modification pathologies in the Internet''. EASE returns a Steiner graph consisting of a single node that includes the paper name itself. AKSDV could find all of the information about this paper; $Q_{11}$ wants to look for Abdul Rahman's friends who have bought BURRDA and given a perfect rating. For EASE, it returns answers mainly about "Abdul Rahman". But AKSDV could return the relevant information which users want to find.

In these three tables, the ``$min\_threshold$'' decides which co-location compounds will be regarded as compounds $\kappa$ = \{$w_1$, ..., $w_p$\}. Each column corresponds to the performance of keyword searches when assigned a value to $min\_threshold$. For example, in Table \ref{tab:prfDBLP}, the second column illustrates that when we set $min\_threshold$ = 0, the values of average precision, recall, and F-measure of AKSDV on the DBLP data set are 0.830, 0.867, and 0.834, respectively. In general, our method of identifying compounds works well on these datasets for improving query performance.

Column ``non'' means that we assign an average weight ($\sigma_i^2 = 1/p$, $||\kappa||$ = $p$) to each $w_i$ in the compound $\kappa$ instead of considering which word will have more contributions to the compound (without using co-location weight).
Column ``0'' and ``non'' regard all the co-location compounds as compounds. The difference between them is whether using co-location weight when constructing compounds. Table~\ref{tab:prfDBLP} and Table~\ref{tab:prfUniBench} demonstrate our co-location weight is better than the average weight method on the DBLP and UniBench dataset. In Table~\ref{tab:prfIMDB}, the performance of column ``0'' is little less than the column ``non''s.


\begin{figure}[!tbp]
\includegraphics[height=3.7cm,width=13cm]{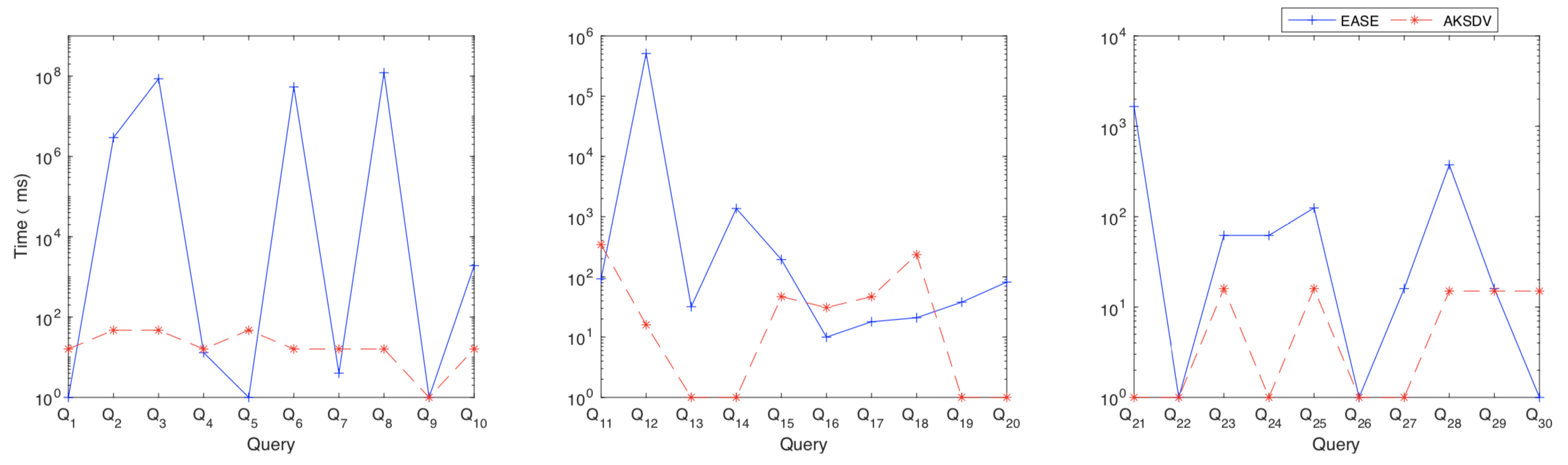}
\caption{Execution time of $Q_i$}
\label{fig:TotalTime}
\end{figure}

\begin{figure*}[!tbp]
\centering
\begin{tikzpicture}
\scriptsize
\begin{groupplot}[
                width=5.5cm,  
        		height=3.7cm,
        		group style={group size=3 by 3, horizontal sep=1.5em, vertical sep=4em}]
        		
\nextgroupplot[
        legend pos=north west,
        legend style={font=\tiny},
		title= (1) DBLP dataset,
		xlabel=Loading Percentage,
		ylabel=$Average\ Time (ms)$,
		ymin = 0,
		ylabel style={at={(-0.16,0.45)}, anchor=north},
        xtick={40,60,80,100},
		xticklabels={40\%, 60\%, 80\%, 100\%}
		]
\addplot[color=red,mark=diamond,thick,dashed] coordinates {
(40,4.7)
(60,9.5)
(80,18.7)
(100,23.7)};
\addlegendentry{AKSDV}

\nextgroupplot[
        legend pos=north west,
        legend style={font=\tiny},
		title= (2) UniBench dataset,
		xlabel=Loading Percentage,
		ymin = 0,
		ymax = 80,
		ylabel style={at={(-0.11,0.4)}, anchor=north},
        xtick={40,60,80,100},
		xticklabels={40\%, 60\%, 80\%, 100\%}
		]
\addplot[color=red,mark=diamond,thick,dashed] coordinates {
(40,15.5)
(60,21.9)
(80,38.9)
(100,71.8)};
\addlegendentry{AKSDV};

\nextgroupplot[
        legend pos=north west,
        legend style={font=\tiny},
		title= (3) IMDB dataset,
		xlabel=Loading Percentage,
		ymin = 0,
		ylabel style={at={(-0.11,0.4)}, anchor=north},
        xtick={40,60,80,100},
		xticklabels={40\%, 60\%, 80\%, 100\%}
		]
\addplot[color=red,mark=diamond,thick,dashed] coordinates {
(40,4.7)
(60,4.7)
(80,6.2)
(100,7.7)};
\addlegendentry{AKSDV}

\end{groupplot} 
\end{tikzpicture}
\caption{Scalability}
\label{fig:scalability}
\end{figure*}
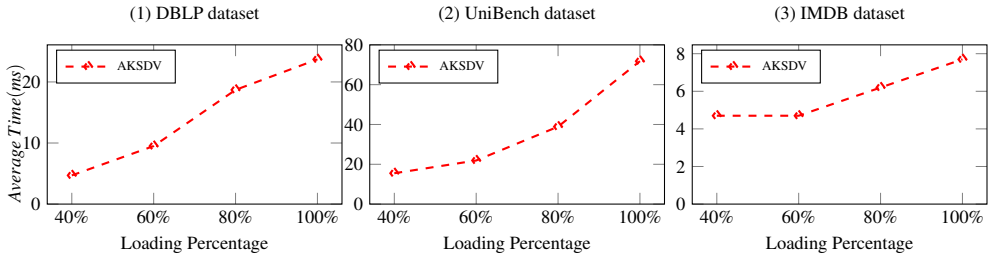




\begin{figure*}[!tbp]
\centering
\tiny
\begin{tikzpicture}
 \pgfplotstableread{dataDBLP.csv}{\loadeddataone}
 \pgfplotstableread{dataUniBench.csv}{\loadeddatatwo}
 \pgfplotstableread{dataFilm.csv}{\loadeddatathree}
 \begin{groupplot}[
     group style={group size=3 by 3, horizontal sep=2em, vertical sep=7em},
     height = 4cm,
     width = 5.5cm,
     ymin = 0,
     xtick = {1,...,25},
     xticklabels = {1,...,25}, 
    ]
    
    
     \nextgroupplot[
        xmin = 1, 
        ymax = 310,
		title= (1) DBLP ($n$: 283 253),
		xlabel= $C'_{denSys}$,
		ylabel=The number of eigenvalues $h$,
		ymin = 0,
		ylabel style={at={(-0.17,0.5)},anchor=north},
		xticklabels from table = {\loadeddataone}{number}, xtick = {1,...,25},
		]
        \addplot[color=red,mark=diamond,thick] table [x = number, y = value] {\loadeddataone};
        \addplot[color=blue,thick,dashed] coordinates {
        (1,200)
        (25,200)};

    
     \nextgroupplot[
        xmin = 1, 
        ymax = 310,
		title= (2) UniBench ($n$: 5666),
		xlabel= $C'_{denSys}$,
		xticklabels from table = {\loadeddatatwo}{number}, xtick = {1,...,25},
		]
        \addplot[color=red,mark=diamond,thick] table [x = number, y = value] {\loadeddatatwo};
        \addplot[color=blue,thick,dashed] coordinates {
        (1,200)
        (25,200)};
    
    
    \nextgroupplot[
        xmin = 1, 
        ymax = 310,
		title= (3) IMDB ($n$: 249 353),
		xlabel= $C'_{denSys}$,
		ymin = 0,
		ylabel style={at={(-0.1,0.5)},anchor=north},
		xticklabels from table = {\loadeddatathree}{number}, xtick = {1,...,25},
		]
        \addplot[color=red,mark=diamond,thick] table [x = number, y = value] {\loadeddatathree};
        \addplot[color=blue,thick,dashed] coordinates {
        (1,200)
        (25,200)};
    \end{groupplot}

\end{tikzpicture}
\caption{The change in value of $h$ on different datasets}
\label{fig:changeOfp}
\end{figure*}
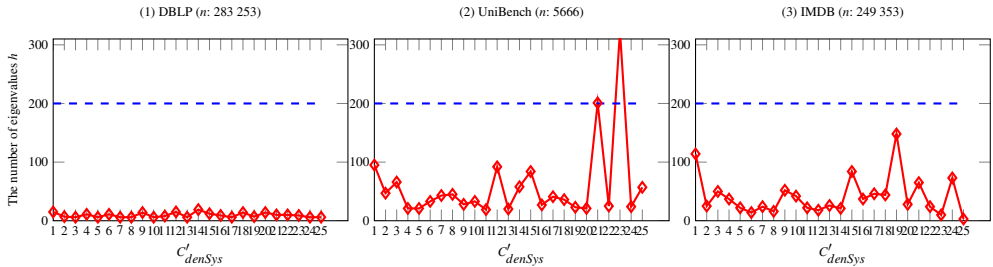

\subsection{Time and Scalability Analysis}
In this part of the experiments, firstly, we analyze the time cost of AKSDV and EASE. The reported time cost values are collected by executing each query several times to take the median value. Fig.~\ref{fig:TotalTime} gives executing time of AKSDV ($min\_threshold$ = 0.6) and EASE. 
In  Fig.~\ref{fig:TotalTime}, there is a different fluctuation in the scales of time about different queries, which is caused by the queries of complexity involving different data models. But in general, comparing with EASE, AKSDV has a good time performance.

Now, we focus on the scalability of AKSDV. To test the scalability, we load different proportions of metadata into our framework, respectively. And we let the correct results increase as loading more data for each query. Then, we perform all queries in table \ref{tab:queries} on these datasets having different sizes to get a median time after executing several times. Finally, we calculate the average time on these datasets, respectively. Fig.~\ref{fig:scalability} (1)-(3) show the results of experiments on the DBLP, Unibench, and IMDB dataset. In general, AKSDV has a good scalability performance on these datasets. The UniBench dataset, due to existing mass joins as the loading percentage increases, shows the query time increases faster than others. But generally, it is within an acceptable range.

\subsection{The Dimension of Density Vectors Analysis}
Unlike the word embedding in the machine learning community, which needs a toolkit using lots of time and samples to train vector space models, we only need to extend the events' dimension. And elementary events still keep disjoint feature. Although the dimensions of events have increased, the query's cost still depends on the $h$ (the number of eigenvalues of density matrices). The Fig.~\ref{fig:changeOfp} shows the change in values of $h$ on the candidate statement set $C'_{eigen}$, where $C'_{eigen} \subset C_{eigen}$ and $C_{eigen}$ is made of candidate statements about all the queries on different datasets in Table~\ref{tab:queries}, and $C'_{eigen}$ is made of selected 25 candidate answers from the corresponding $C_{eigen}$ randomly. In Fig.~\ref{fig:changeOfp}, we can see the values of $h$ are much less than the word space $n$, even less than 200 in most situations. It guarantees the complexity of our framework at the appropriate level.



\section{Related works}
Keyword search has been well known as a user-friendly way of satisfying users' information needs with few keywords in diverse fields such as Information Retrieval (IR) and database.
Unfortunately, finding a fundamental axiomatic formulation of the IR field has proven tricky, maybe because of the intrinsic role humans play in the process \cite{2018arXiv180905685A}. However, due to information having proven considerably easier to capture mathematically than ``meaning'', researchers are investigating how to apply quantum theory to attack the various IR challenges, which can be traced back to \cite{van1989towards}. Piwowarski et al. \cite{piwowarski2010can} used the probabilistic formalism of quantum theory to build a principled interactive IR framework. Frommholz et al. \cite{frommholz2010supporting} utilized poly representation and quantum measurement \cite{wang2010tensor} to do keyword searches in a quantum-inspired interactive framework.

Considering the power of quantum-inspired framework and the similarity of keyword searches between databases and IR, we attempt to use the quantum-inspired method to support keyword searches on multi-model databases. Hence our research work also conforms to the current trend of seamlessly integrating database and information retrieval \cite{weikum2007db}. The keyword search on the database is particularly appealing. From relation, XML to graph databases, there are already many exciting proposals in the scientific literature \cite{kargar2015meaningful,hristidis2003keyword,guo2003xrank,he2007blinks}. However, most of the existing keyword search works are designed for specific data models. Through review literature, the only complete relevant current work is EASE \cite{li2011effective}.
They returned r-radius Steiner graphs as results to users for answering keyword searches on heterogeneous data.
In addition, there is another work \cite{lin2017towards}, which presented an architecture to support keyword searches over diverse data sources. Due to lacking complete performance analysis, we temporarily do not compare it in this paper.


\section{Conclusion}
This paper has proposed using the quantum probability mechanism to solve the keyword search problem on multi-model databases. To reduce complexity and improve performance, we introduced new representations of events and the density vector concept. We also used the spatial pattern to help improve query accuracy and used PCA to support the framework to work well further. 
Finally, extensive experiments have demonstrated the superiority of our methods over state-of-the-art works.

\subsubsection*{Acknowledgements.} The work is partially supported by the China Scholarship Council and the Academy of Finland project (No. 310321). We would also like to thank all the reviewers for their valuable comments and helpful suggestions.

%
%
%
%
%
%






\end{document}